\documentstyle[twocolumn,aps,floats]{revtex}
\begin{document}
\tighten
\draft
\newcommand{\ds}{\displaystyle}
\newcommand{\be}{\begin{equation}}
\newcommand{\en}{\end{equation}}
\newcommand{\bea}{\begin{eqnarray}}
\newcommand{\ena}{\end{eqnarray}}
\topmargin 0cm

\title{Comment on ``Viscous cosmology in the Kasner metric".}
\author{Mauricio Cataldo$^{\,\,a \, b}$
{\thanks{E-mail address: mcataldo@alihuen.ciencias.ubiobio.cl}}
and Sergio del Campo $^{c}$
\thanks{E-mail address: sdelcamp@ucv.cl}}
\address{$^a$
Departamento de F\'\i sica, Facultad de Ciencias,
Universidad del B\'\i o-B\'\i o, Avda. Collao 1202, Casilla 5-C, Concepci\'on,
Chile.
\\
$^c$ Instituto de F\'\i sica, Facultad de Ciencias B\'asicas y matem\'aticas,
Universidad Cat\'olica de Valpara\'\i so,
Avenida Brasil 2950, Valpara\'\i so, Chile.
}
\maketitle
\begin{abstract}
{\bf {Abstract:}} We show in this comment that in an anisotropic
Bianchi type I model of the Kasner form, it is not possible to
describe the growth of entropy, if we want to keep the
thermodynamics together with the dominant energy conditions. This
consequence disagrees with the results obtained by Brevik and
Pettersen [Phys. Rev. D {\bf 56}, 3322 (1997)].

\vspace{0.5cm}

PACS number(s): {98.80.Hw, 98.80.Bp}

\end{abstract}

\smallskip\

Brevik and Pettersen~\cite{Brevik} have studied the consequences
that a Bianchi type I metric of the Kasner form
\begin{eqnarray}
\label{Metric}
  \ds ds^2= - dt^2 + t^{2 p_1} dx^2+ t^{2 p_2} dy^2+
t^{2 p_3} dz^2
\end{eqnarray}
occurs for the equation of state for the  cosmic fluid,
characterized by a shear viscosity $\eta$ and a bulk viscosity
$\xi$. In their work, they concluded that for a viscous fluid,
with $\eta \neq 0$ and $\xi \neq 0$, and from Einstein equations,
the requirement that the three Kasner parameters $p_{i}$ ($i =
1,2,3$) be constant implies that $\eta \sim 1/t$ and $\xi \sim
1/t$.

 From their equation (29) it is obtained an explicit expression for
the shear viscosity given by
\begin{eqnarray}
\label{expresion 2}
\ds \eta = \frac{1}{2 \kappa t} (1-S),
\end{eqnarray}
where $\kappa= 8 \pi G$ and $\ds S= \sum _{i = 1}^{3} p_{i}$. It
could be shown from thermodynamics that we should impose  the
condition $\eta \geq 0$~\cite{MiThWe}. Therefore, we see from
expression~(\ref{expresion 2}) that we must require that
\begin{eqnarray}
S-1 \leq 0.
\end{eqnarray}
On the other hand, the entropy production, in an anisotropic Kasner type
universe, becomes
\begin{eqnarray}
\label{expresion 3}
\dot{\sigma} \approx \frac{2 S^2}{n k_{_B} T t^2} \eta A,
\end{eqnarray}
where $\ds A= 1/3 \sum_{i = 1}^3 (1-H_i/H)^2=3Q/S^2-1 \geq 0$,
$\ds Q= \sum_{i = 1}^{3} p^2_{i}$, $n$ is the baryon number
density, $k_{_B}$ is the Boltzmann constant and $T$ the
temperature.

In expression~(\ref{expresion 3}) we have restricted to the case
in which the shear viscosity $\eta$ is vastly greater than the
bulk viscosity $\xi$, as was considered in Ref.~\cite{Brevik}.

Thus, following the result obtained by Brevik and Pettersen, we conclude
that in an anisotropic Kasner type model, the  parameters entering in the
metric have to satisfy the bound $S \leq 1$, in order to deal with an
appropriated physical model.

On the other hand, if we require the model to satisfy the dominant
energy conditions, specified by $- \rho \leq  P_j \leq
\rho$~\cite{Hawking}, where $\rho$ is the energy density and $P_j$
(with $j = x,y,z$) are the effective momenta related to the
corresponding coordinate axis, we could show that the shear
viscosity $\eta$ necessarily, in this sort of model, becomes
negative, since these dominant energy conditions imply that $S
\geq 1$, and not $S \leq 1$, as was specified by Brevik and
Pettersen.

To see this, let us write the dominant energy condition explicitly
in terms of the parameters that enter in the metric
(\ref{Metric}), i.e. $p_1, p_2$ and $p_3$.

Einstein field equations can be written in comoving coordinates as
(with $\kappa = 8 \pi G = 1$)
\be
\label{00} \ds
\frac{\dot{a}}{a}\frac{\dot{b}}{b}+\frac{\dot{a}}{a}\frac{\dot{c}}{c}+
\frac{\dot{b}}{b}\frac{\dot{c}}{c}= \rho,
\en
\be
\label{11} \ds
\frac{\ddot{b}}{b}+\frac{\ddot{c}}{c}+\frac{\dot{b}}{b}\frac{\dot{c}}{c}
= -P_x,
\en
\be
\label{22} \ds
\frac{\ddot{a}}{a}+\frac{\ddot{c}}{c}+\frac{\dot{a}}{a}\frac{\dot{c}}{c}
= -P_y
\en
and
\be
\label{33} \ds
\frac{\ddot{a}}{a}+\frac{\ddot{b}}{b}+\frac{\dot{a}}{a}\frac{\dot{b}}{b}
= -P_z,
\en
where $a$, $b$ and $c$ are the anisotropic expansion factors. From
the metric (\ref{Metric})  they are given by $a=t^{p_1}$,
$b=t^{p_2}$ and $c=t^{p_3}$. Thus, the Einstein´s field equations
(\ref{00})-(\ref{33}) reduce to
\be
\label{ro} \ds \rho = \frac{p_1 p_2 + p_1 p_3 + p_2 p_3}{t^2},
\en
\be
\label{P1} \ds P_x = -\frac{p_2^2 + p_3^2 -p_2 - p_3 + p_2
p_3}{t^2},
\en
\be
\label{P2} \ds P_y = -\frac{p_1^2 + p_3^2 -p_1 - p_3 + p_1
p_3}{t^2}
\en
and
\be
\label{P3} \ds P_z = -\frac{p_1^2 + p_2^2 -p_1 - p_2 + p_1
p_2}{t^2}.
\en
Here, as was mentioned above, $P_j$, with $j = x,y,z$, represent
the effective momenta in the corresponding coordinate axis. Note
that either $\rho$ and $P_j $ (with $j=x,y,z$) scale as $t^{-2}$.
Thus, the dominant energy conditions will give some specific
relations between the Kasner parameters $p_i$.

The conditions $P_j \leq \rho$, with $j = x,y,z$, yield to three
inequations given by
\be
\label{ine1} \ds (S-p_1)(S -1)\geq 0,
\en
  \be
\label{ine2} \ds (S-p_2)(S -1)\geq 0
\en
and
\be
\label{ine3} \ds (S-p_3)(S -1)\geq 0,
\en
which, after adding them, reduced to just one inequation given by
\begin{eqnarray}
\label{expresion 5}
2 S (S-1) \geq 0.
\end{eqnarray}
In a similar way, from $P_j \geq - \rho$ we get the inequations
\begin{eqnarray}
S(1+p_1)-p_1 \geq Q,
\end{eqnarray}
\begin{eqnarray}
S(1+p_2)-p_2 \geq Q,
\end{eqnarray}
and
\begin{eqnarray}
S(1+p_3)-p_3 \geq Q.
\end{eqnarray}
After adding them we get
\begin{eqnarray}
\label{expresion7}
S \geq \frac{A S^2}{2}.
\end{eqnarray}

 From expression~(\ref{expresion7}), we see  that $S \geq 0$, since
by definition $A \geq 0$. With this condition on $S$, we obtain
from expression~(\ref{expresion 5}) that necessarily $S$ should be
greater than one. This means that a Bianchi type I metric of the
Kasner form, always give rise to a negative shear viscosity, where
expression (\ref{expresion 2}) applies. From this result, and from
expression~(\ref{expresion 3}), we observe that we are finished
with the unfavored situation in which $\dot{\sigma} \leq 0$,
giving the meaning that the entropy in this sort of universe
decreases instead of increasing.

In conclusion, we have shown in this comment that it is not
possible to describe the growth of entropy in the universe for a
viscous anisotropic Bianchi type I metric of the Kasner form, if
we want to keep the thermodynamic conditions together with the
dominant energy conditions.

MC and SdC were supported by COMICION NACIONAL DE CIENCIAS Y
TECNOLOGIA through Grants FONDECYT N$^0$ 1990601 and N$^0$
1971157. Also MC was supported by Direcci\'{o}n de Promoci\'{o}n y
Desa-rrollo de la Universidad del B\'{\i}o-B\'{\i}o, and SdC was
supported from UCV-DGIP 123.744/99.

\end{document}